\documentclass[aps, prd, reprint, showpacs, superscriptaddress, twocolumn, nofootinbib]{revtex4-1}    

\usepackage{graphicx}
\usepackage{dcolumn}
\usepackage{bm}
\usepackage{natbib}
\usepackage{amsmath}

\renewcommand{\v}[1]{\ensuremath{{\mathbf{#1}}}}
\newcommand{\ketnu}[1]{\ensuremath{\vert \nu_{#1} \rangle}}
\newcommand{\nut}[1]{\ensuremath{\nu_{#1}}}
\newcommand{\nutbar}[1]{\ensuremath{\bar\nu_{#1}}}

\begin{document}

\title{Quantum Kinetic Equilibrium}

\author{Chad T.\ Kishimoto}
\email{ckishimoto@sandiego.edu}
\affiliation{Department of Physics and Biophysics, University of San Diego, San Diego, CA 92110}
\affiliation{Center for Astrophysics and Space Sciences, University of California, San Diego, La Jolla, CA 92093}
\author{Heather Hodlin}
\affiliation{Department of Physics and Biophysics, University of San Diego, San Diego, CA 92110}
\author{Olexiy Dvornikov}
\affiliation{Department of Physics and Astronomy, University of Hawai`i at M\={a}noa, Honolulu, HI 96822}

\date{\today}

\begin{abstract}
We solved the Quantum Kinetic Equations (QKEs) for an active-sterile neutrino system in the early universe.  While on the surface this may seem to be an overly simplistic system, other linear two-state systems can be mapped onto the active-sterile system.  In the early universe, we find that solutions to the QKEs are well described by an adiabatic approximation where the off-diagonal terms of the density operator are constant on short (oscillation and/or scattering) timescales, but may slowly evolve on long (expansion) timescales.  This ``quantum kinetic equilibrium'' attains as the quantum development of phase balances with the kinetic destruction of phase.  In this work, we introduce and assess this equilibrium ansatz as the neutrino states evolve in the early universe with a non-zero lepton number, engendering level crossings that result in scattering-induced active-sterile neutrino transformation.
\end{abstract}

\maketitle

\section{Introduction}

Studying the evolution of quantum particles that scatter with their environment is important in a wide array of applications, from quantum computing to neutrino astrophysics.  Neutrino states are an archetype quantum system where the non-equivalence of their energy and interaction (``flavor'') eigenstates ensure the evolution of quantum phase between the neutrino flavor states.  Furthermore, in hot and dense astrophysical environments, neutrino scattering (``kinetics'') influences this development of quantum phase.  The Quantum Kinetic Equations (QKEs) self-consistently solve for this evolution that involves both the development and destruction of quantum phase \cite{sr93, mt94, sb05,vfc14,cfv15}. 

In astrophysical environments, the quantum kinetic evolution of neutrinos is imprinted on a variety of phenomena where they play a principal role in energy/entropy transport and nucleosynthesis.  In the early universe, this quantum kinetic evolution directly affects the energy density in neutrinos (as parameterized by the derived CMB parameter $N_{\rm eff}$) and the yield of big bang nucleosynthesis \cite{man05,dp16,fpv20} and may play a role in other beyond-Standard Model processes. 
Elsewhere, near the proto-neutron star in the core of a supernova, the quantum kinetic evolution of neutrinos will affect the transport of energy and entropy from the proto-neutron star to the envelope and influences the synthesis of $r$-process materials in the supernova \cite{dk09,dfq10,mfm14,vm16,saw16,dms17}.  In these environments, neutrinos are created in prodigious numbers where it is hot and dense and these large fluxes subsequently diffuse toward cooler and less dense regions.  Large neutrino fluxes mean that neutrino-neutrino interactions make their quantum kinetic evolution nonlinear.

In this work, we solve the QKEs for the {\it linear} evolution of a two-state, active-sterile neutrino system in the early universe with a non-zero lepton number.  This represents the same physical basis as work done to self-consistently calculate the scattering-induced decoherent production of sterile neutrino dark matter \cite{vcah16}.  However, Ref.\ \cite{vcah16} presents a non-linear Boltzmann-like approach to the problem, while this work is a significantly more computationally expensive linear quantum kinetic approach.  It has been shown  that the Boltzmann-like approach to producing sterile neutrino dark matter is consistent with solutions to the QKEs \cite{kf08}.  The goal of this work is to extend this analysis to the evolution of the quantum coherences to understand the characteristics of quantum kinetic evolution in the early universe, rather than self-consistently solve the fully non-linear problem, which will be left to future work.

Linearity in the QKEs is achieved by neglecting the non-linear feedback of the quantum kinetic evolution of neutrinos in the neutrino-neutrino interaction.  While the non-linearity is beyond the scope of this work, we will argue that since this non-linear feedback is significant on universal expansion time scales, which is much longer than the oscillation and scattering time scales, the general characteristics of the quantum kinetic evolution of the neutrino states would be unchanged by such slow feedback.  

We simulated the quantum kinetic evolution of an active-sterile neutrino system ($\nu_e$-$\nu_s$) with a 7.1 keV dark-matter-candidate sterile neutrino and mixing angle $6 \times 10^{-11}$, consistent with X-ray observations attributed to the decay of sterile neutrino dark matter \cite{bulbul14}. 
To mimic the resonances that produce a significant quantity of dark matter, we introduce a net lepton number (excess of neutrinos over anti-neutrinos relative to the photon number) in each neutrino species, $L_{\nu_e} = L_{\nu_\mu} = L_{\nu_\tau} = 9 \times 10^{-4}$.  As in Ref.\ \cite{kf08}, we neglect the changing relativistic degrees of freedom that result from the annihilation of Standard Model particles as the temperature decreases with the expanding universe and the loss of quark degrees of freedom from the QCD transition.  Just as before, these changes occur on expansion time scales and should not affect the conclusions of this work.  

\begin{figure}[t]
    \includegraphics[width=0.8\columnwidth]{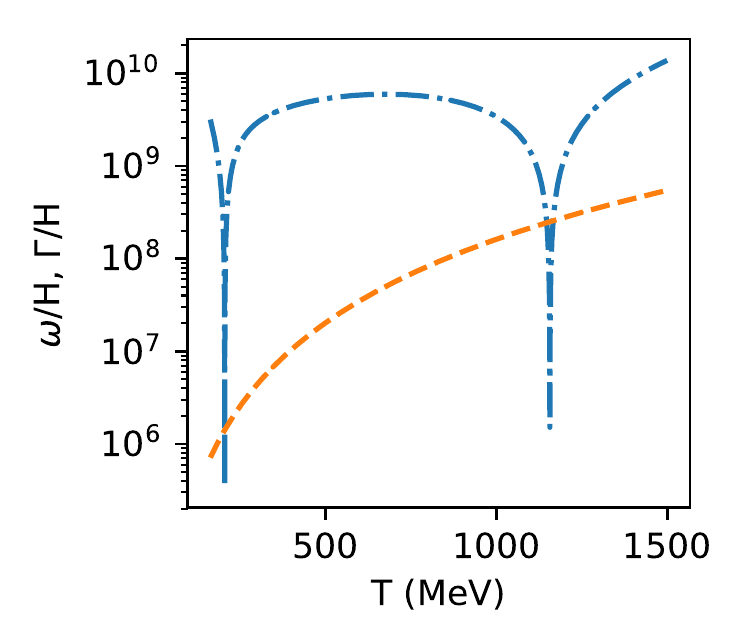}
    \caption{The ratio of the oscillation rate to the expansion rate $\omega / H$ (dashed curve) and the ratio of the scattering rate  to the expansion rate $\Gamma / H$ (dot-dashed curve).}
    \label{fig:rates}
\end{figure}

The quantum kinetic evolution of the neutrino system is governed by three rates:  the in-medium neutrino oscillation rate, $\omega$, the active neutrino scattering rate with the background, $\Gamma$, and the expansion rate, $H$.  Figure \ref{fig:rates} shows that the oscillation and scattering rates are always much faster than the expansion rate in our calculation.  

The oscillation rate shows the existence of two resonances where the in-medium mixing is maximal and the oscillation rate rapidly changes over many orders of magnitude.  Sterile neutrino dark matter is primarily produced at these resonances, where their otherwise resonant production is suppressed by the quantum Zeno effect.  It has been shown that the oft-used quantum Zeno ansatz \cite{afp,vcah16} 
for the de-coherent active-sterile transformation rate is consistent with solutions to the QKEs, at least at the percent level \cite{kf08,johns19}.  This work investigates the evolution of the quantum coherences between the states, as well as the occupation probabilities.

We find that in this environment (with the oscillation and scattering rates much faster than the expansion rate, $\omega, \Gamma \gg H$), solutions to the QKEs are well-described by an adiabatic approximation where the off-diagonal terms of the density operator are constant on short (oscillation and/or scattering) timescales, but may slowly evolve on long (expansion) timescales.  We dub this effect, ``quantum kinetic equilibrium.''  

We set $\hbar = c = k_B = 1$ in this work.  In Section \ref{sec:solve} we introduce the QKEs and discuss the relevant approximations taken and our solution techniques.  Section \ref{sec:ringing} introduces the initial conditions, and how solutions to the QKEs ``ring'' toward a robust quantum kinetic equilibrium, then in Section \ref{sec:qkeq} we discuss how this equilibrium evolves as the universe expands and the slight deviations of solutions to the QKEs from this equilibrium.  Finally, we discuss the caveats to this calculation and present conclusions in Section \ref{sec:conc}.

\section{Solving the QKEs} \label{sec:solve}

The QKEs are a set of quantum master equations for the density operator, 
\[  f = \left( \begin{array}{cc} f_{ee} & f_{es} \\ f_{es}^* & f_{ss} \end{array} \right), \]
where the diagonal elements are the distribution functions of both flavor states and the off-diagonal term encodes information about the quantum coherences.  The QKEs can be written as
\begin{equation}
    \mathcal{D} f = -i \left[ \mathcal{H}, f \right] + \mathcal{C} [ f ] ,
\end{equation}
where $\mathcal{D}$ is the convective derivative, $\mathcal{H}$ is the Hamiltonian, and $\mathcal{C} [f]$ the collision operator, which is a functional of the density operator.

In this work, we will consider the simplest possible two-state system, one with an active (\nut{e}) and sterile (\nut{s}) state, related by a vacuum mixing angle, $\theta$:
\begin{eqnarray}
\ketnu{e}  = &\cos \theta \ketnu{1} + \sin \theta \ketnu{2} \nonumber \\
\ketnu{s}  = &- \sin \theta \ketnu{1} + \cos \theta \ketnu{2} , \label{eq:nu_mix}
\end{eqnarray}
where \nut{1,2} represent the low and high mass states of the vacuum Hamiltonian, respectively.  The active state has forward scattering potential $V$ and scattering rate $\Gamma$ with the background universe, while the sterile state has none.  Both $V$ and $\Gamma$ are time dependent because of universal expansion.

The collision integral can be written as
\begin{equation}
    \mathcal{C} [f] = - \Gamma \left( \begin{array}{cc}  \Delta f_e & 0 \\ 0 & 0 \end{array} \right) - \frac{\Gamma}{2} \left( \begin{array}{cc} 0 & f_{es}  \\ f_{es}^* & 0 \end{array} \right) ,
\end{equation}
where $\Delta f_e = f_{ee} - f_e^{\rm (eq)}$ is the difference between the $\nut{e}$ distribution function, $f_{ee}$, and the equilibrium, Fermi-Dirac distribution function, $f_e^{\rm (eq)}$.  This is the approximate form of the collision operator  when the \nut{e} are near equilibrium, which is likely appropriate because $\Gamma \gg H$ in our regime of interest.

As is commonly done, we decompose the density operator as
\begin{equation}
    f = \frac{1}{2} P_0 \left( 1 + \mathbf{P} \cdot \boldsymbol{\sigma} \right) ,
\end{equation}
where $1$ is the identity, $\boldsymbol{\sigma}$, are the Pauli spin operator, $P_0$ is a normalization proportional to the total number of neutrinos, and $\mathbf{P}$ is a polarization vector.  The QKEs are the equations of motion for $P_0$ and $\mathbf{P}$ \cite{mt94,sr93}
\begin{align}
    \mathcal{D} \,\v{P} & = \v{V} \times \v{P} + (1 - P_z) \left( \mathcal{D} \ln P_0 \right) \,\hat{\mathbf{z}} \nonumber \\
    & \qquad \qquad - \left( \frac{1}{2} \Gamma + \mathcal{D} \ln P_0 \right) \, (P_x \, \hat{\mathbf{x}} + P_y \, \hat{\mathbf{y}} ) \label{eq:qke_P} \\
    \mathcal{D} \, P_0 & = \Gamma \left[ 1 - \frac{1}{2} P_0 ( 1 + P_z )  \right] = R, \label{eq:qke_P0}
\end{align}
using the same conventions as in Ref.\ \cite{kf08}.  The vector 
\begin{equation}
    \v{V} = \frac{m_s^2}{2 p} ( \sin 2 \theta \, \hat{\mathbf{x}} - \cos 2 \theta \, \hat{\mathbf{z}} ) + V \, \hat{\mathbf{z}} ,
\end{equation}
where $p$ is the neutrino momentum and $m_s$ is the sterile neutrino mass (with the approximation that the sterile mass is much larger than the active neutrino masses), can be interpreted as causing coherent quantum mechanical evolution.  The repopulation function, $R$, efficiently repopulates \nut{e} so that it remains in thermal and chemical equilibrium with the environment.

It will be useful for us to describe the ``weak isospin'' space in which \v{P} and \v{V} operate.  The polarization vector, \v{P}, contains information about the neutrino state.  The z-projection of \v{P}, $P_z$, is the difference of the occupation probabilities of \nut{e} minus \nut{s}, which means if $\v{P} = \hat{\mathbf{z}}$, then there are only \nut{e} and no \nut{s}.  The x- and y-components of \v{P} represent quantum mechanical coherences between the two flavor states.  The Hamiltonian vector, \v{V}, points in the direction of the high energy eigenstate and its magnitude is equal to the coherent neutrino oscillation frequency, which is equal to the difference in the energy eigenvalues.  It should be observed that in this case \v{V} is not parallel to the z-axis, which means that the energy eigenstates are never coincident with the flavor eigenstates, which generates the coherent evolution of the neutrino states.

In the absence of collisions (the limit as $\Gamma \rightarrow 0$), solutions to the QKEs are simply those of the coherent evolution.  \v{P} precesses around \v{V} with oscillation frequency $\omega = \vert \v{V} \vert$, which corresponds to an oscillation in the probability of measuring the neutrino in either flavor state (z-projection of \v{P}) and the time evolution of the coherences between flavor states (the off-diagonal components of the density operator, represented by the x- and y-components of \v{P}).

The effect of collisions (the other terms in Eq.\ (\ref{eq:qke_P}), which are proportional to $\Gamma$) is to reduce the length of $\vert \v{P} \vert$ as the coherences are damped toward zero.  In the limit that $\Gamma \rightarrow \infty$, the coherences are zero (\mbox{$P_x = P_y = 0$,} the off-diagonal terms in the density operator are zero), preventing oscillation between the flavor states (quantum Zeno effect) and forcing the neutrinos into thermal equilibrium.  This mimics the collapse of the wave function into one of the flavor states.  A finite scattering rate allows for non-zero coherences to form between the flavor states.

In the early universe, the forward scattering potential for neutrinos can be written as the sum of a finite density potential arising from non-zero lepton numbers (first line below) and a thermal potential from thermal populations of leptons (second line):
\begin{align}
V   =  & \frac{2 \sqrt{2} \zeta (3)}{\pi^2} G_F T^3 (2 L_{\nut{e}} + L_{\nut{\mu}} + L_{\nut{\tau}} )  \label{eq:V} \\ 
   & ~~ - \frac{8 \sqrt{2} G_F p}{3 m_W^2} \left[ \rho_{e^-} + \rho_{e^+} + \left( \frac{m_W}{m_Z} \right)^2 ( \rho_{\nut{e}} + \rho_{\nutbar{e}} ) \right] , \nonumber
\end{align}
where $G_F$ is the Fermi constant, $m_W$ and $m_Z$ are the $W^\pm$ and $Z^0$ boson masses, and the $\rho$ are the energy density in electrons, positrons, \nut{e}, and \nutbar{e}, respectively.  We take the lepton numbers to be constant and the energy densities to be their thermodynamic values.  

It should be noted that a fully self-consistent solution to the QKEs would require the forward scattering potential to be a matrix where the lepton numbers and neutrino energy densities should be integrals over the density matrices of neutrinos and of an entire spectrum of momenta as well as the density matrices of anti-neutrinos.
There are two primary differences between our calculation and a fully self-consistent solution:  1) we treat the lepton numbers as constant, while the self-consistent solution will have an evolving lepton number as \nut{e} are converted to \nut{s}; and 2) we neglect the off-diagonal potential that results from non-zero off-diagonal terms in the density operators that would create x- and y-components of \v{V}.  Lepton numbers will change as \nut{e} transform into \nut{s}, but we expect these to change on the slow timescale of the expanding universe and have negligible effect on the conclusions of this work.  The off-diagonal terms of the density operator are small compared to the diagonal terms, so the off-diagonal potential should have small effects on the numerical results in this  linear work, but we do not expect to it to affect the conclusions.  However, in a fully self-consistent, non-linear calculation, collective effects may produce new and interesting conclusions.  This remains a fruitful area for further investigation.

\begin{figure*}
    \includegraphics[width=0.8\textwidth]{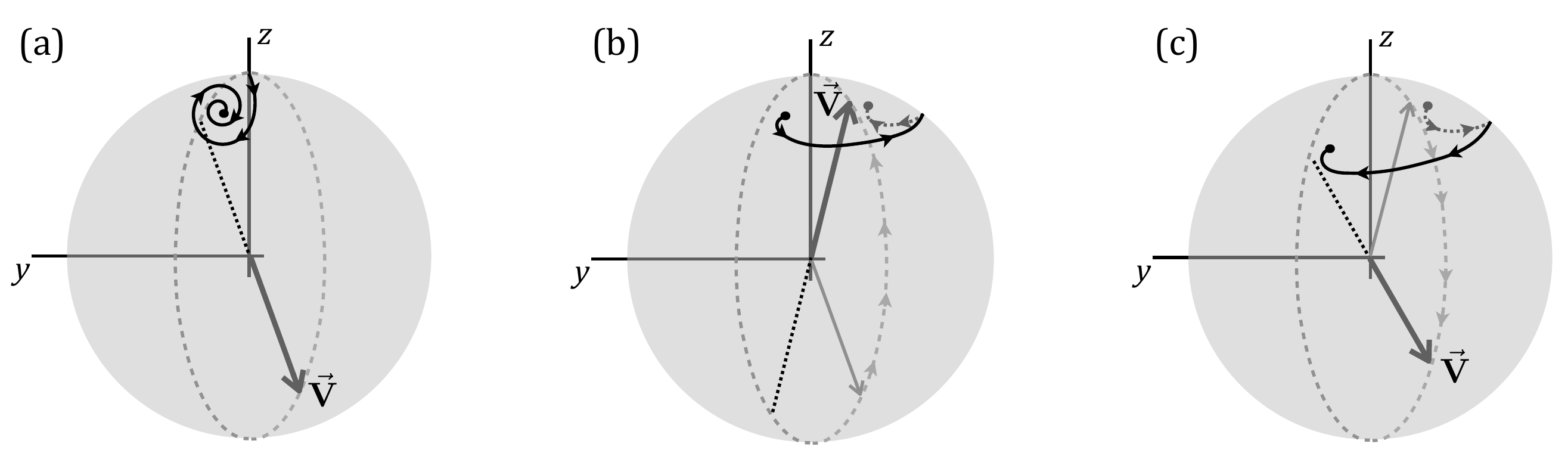}
    \caption{A Bloch sphere representation of the evolution of the polarization vector, shown as the solid-line trajectory.  For reference, the direction of the  Hamiltonian vector, \v{V}, is included.  Figure (a) shows the initial conditions ($\v{P} = \hat{\mathbf{z}}$) ``ringing'' toward a quantum kinetic equilibrium (Section \ref{sec:ringing}).  Figures (b) and (c) show the resonances where \v{V} rapidly flips, the polarization vector tracks the quantum kinetic equilibrium (Section \ref{sec:qkeq}).  Note that in this sketch, angle between the z-axis and both \v{P} and \v{V} are greatly exaggerated for the sake of presentation.}
    \label{fig:QKE_solutions}
\end{figure*}

Expansion of the early universe is described by the scale factor, $a$, in the Friedmann-Lema\^{i}tre-Robertson-Walker metric.  The evolution of the scale factor is governed by the Friedmann equation,
\begin{equation}
    \frac{1}{a} \frac{d a}{d t} = H = \sqrt{\frac{8 \pi \rho}{3 m_{\rm pl}^2}}  = \sqrt{\frac{8 \pi}{3 m_{\rm pl}^2}} \left( \frac{\pi^2}{30} g_* T^4\right)^{1/2} ,
\end{equation}
where $m_{\rm pl}$ is the Planck mass, $\rho$ is the total energy density in the universe, and $g_*$ is the effective number of relativistic degrees of freedom.  In this work we choose a constant value of $g_* = 61.75$, which corresponds to relativistic thermal distributions of quarks (up, down, strange), gluons, leptons (muon, electron), neutrinos and photons.  This is roughly an appropriate value for the early universe before the QCD transition for temperatures $\sim 200 - 1500~{\rm MeV}$.  It should be noted that $g_*$ is not constant, rather it varies by tens of percent in this temperature range \cite{kt90, ls06}.  However, as we have argued before, $g_*$ changes on slow expansion timescales, so we expect no changes to the conclusions reached in this work.

Holding $g_*$ constant means that the plasma temperature evolves as
\begin{equation}
    \frac{1}{T} \frac{dT}{dt} = - \frac{1}{a} \frac{d a}{d t} = - H,
    \label{eq:dTempdt}
\end{equation}
without the reheating of the plasma that occurs from the annihilation of relativistic degrees of freedom as $g_*$ decreases.  Further, when we express the density operator as a function of scaled momentum, $\epsilon = p / T$ the convective derivative in the expanding universe is
\begin{equation}
    \mathcal{D} = \frac{\partial}{\partial t} + \frac{d \epsilon}{d t} \frac{\partial}{\partial \epsilon}  = \frac{\partial}{\partial t} .
\end{equation}

In total, these approximations allow for the QKEs (Eq.\ (\ref{eq:qke_P}) and (\ref{eq:qke_P0})) for a single $\epsilon$ to be a set of four coupled ordinary differential equations that evolve independently of other $\epsilon$.  This effectively linearizes an inherently non-linear and coupled evolution by ignoring the feedback of changing density operators of all epsilon on the forward scattering potential.  Nevertheless, we feel that this work has value in outlining the linear evolution of this system that may serve as a basis for describing the fully non-linear and coupled problem.

We solved the QKEs for neutrinos with $\epsilon = 1$ with the initial condition, $\v{P} = \hat{\mathbf{z}}$ and $P_0 = 1$, which represents the physical scenario of a thermal distribution of $\nut{e}$ and no initial \nut{s}.  To do so, we implemented a Cash-Karp adaptive step size method \cite{numrec} with an overall tolerance level for numerical accuracy of one part in $10^{10}$.  Figure \ref{fig:QKE_solutions} illustrates the evolution of \v{P} on a Bloch sphere.  (The collision operator creates non-unitary evolution, causing $\vert \v{P} \vert$ to deviate from 1.  Nevertheless, the Bloch sphere provides useful visual representation of the evolution of the polarization vector.)  

At the beginning of the simulation, the polarization vector spirals toward a point that is almost anti-parallel to the Hamiltonian vector, but it should be noted that this ``quantum kinetic equilibrium'' is distinct, and not actually anti-parallel to \v{V}.  As the universe expands, the plasma cools and the Hamiltonian vector slowly changes, which in turn causes this equilibrium to slowly change as well.  As discussed above, the parameters chosen in this simulation result in resonances (level crossings where in-medium mixing is maximal) at $T \sim 1200~{\rm MeV}$ and $200~{\rm MeV}$.  At these resonances, the Hamiltonian vector rapidly ``flips'' (as illustrated).  Although this flip is fast compared to the in-medium neutrino oscillation rate, the evolution of the polarization vector tracks this quantum kinetic equilibrium, that balances the influences of both coherent evolution and scattering-induced decoherent evolution.

\section{Initial Conditions and ``Ringing''} \label{sec:ringing}

\begin{figure} 
    \includegraphics[width=\columnwidth]{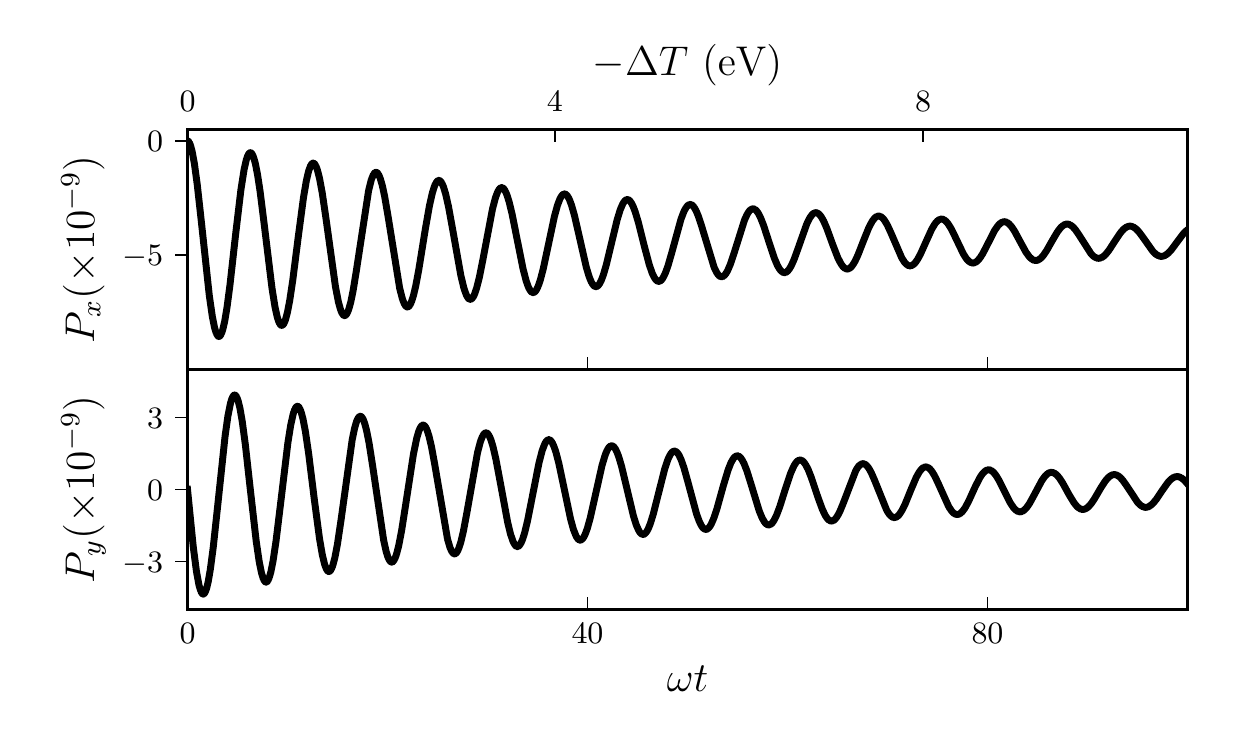}
    \caption{$P_x$ and $P_y$ ring from their initial values toward an equilibrium.  The x-axes are shown equivalently in terms of $\omega t$, the phase of the damped oscillations, and the change in temperature (the negative sign recognizes that the universe cools as time progresses, and it expands).}
    \label{fig:ringing}
\end{figure}

Figure \ref{fig:ringing} shows $P_x$ and $P_y$ ``ringing'' as they evolve from their initial values (zero for both) at $T = 1500~{\rm MeV}$ toward an ``equilibrium'' value.  Both appear to follow damped oscillations with oscillation rate matching the in-medium neutrino oscillation rate, $\omega$, and the damping rate is half the active neutrino scattering rate, $\Gamma / 2$.  Figure \ref{fig:rates} shows that both $\omega$ and $\Gamma$ are large compared to the expansion rate, so this evolution occurs nearly instantaneously in the history of the expanding universe.  In the entire time frame of this ringing phenomenon shown in Figure \ref{fig:ringing}, the temperature of the plasma decreases by roughly 10 eV (compared to $T = 1500~{\rm MeV}$, which corresponds to $\Delta T / T \sim 10^{-8}$).

Along with $P_x$ and $P_y$ ringing toward equilibrium, $P_z$ and $P_0$ do not change from their initial values of 1 (at least within double precision, approximately one part in $10^{16}$).  At equilibrium, we expect to find \mbox{$d P_x / dt = d P_y / dt = 0$}, which leads to equilibrium values, 
\begin{align}
    P_x^{\rm (eq)}  = \frac{V_x V_z}{V_z^2 + \frac{1}{4} \Gamma^2} &\approx - \frac{\omega^2}{\omega^2+\frac{1}{4} \Gamma^2}  \sin 2 \theta_M \nonumber \\
    P_y^{\rm (eq)}  = - \frac{V_x \Gamma}{2 ( V_z^2 + \frac{1}{4} \Gamma^2 )} &\approx -\frac{\omega\Gamma}{2(\omega^2+\frac{1}{4} \Gamma^2)} \sin 2 \theta_M , \label{eq:ringing_eqm}
\end{align}
where $V_x$ and $V_z$ are the x- and z-components of the Hamiltonian vector, and the in-medium effective mixing angle,
\begin{equation}
    \sin 2 \theta_M = \frac{V_x}{\vert \v{V} \vert} ,
\end{equation}
is the sine of the angle between \v{V} and the z-axis in the Bloch sphere representation of Figure \ref{fig:QKE_solutions}.  (By convention, this angle is defined as $2 \theta_M$, which, in vacuum, would be consistent with the vacuum mixing angle in Eq.\ (\ref{eq:nu_mix}).)  At high temperatures, $\sin 2 \theta_M \ll 1$, which leads to the approximations in the equilibrium values, Eq.\ (\ref{eq:ringing_eqm}).

We can obtain a best fit individually for $P_x$ and $P_y$ to a damped oscillator with the general form
\begin{equation}
    P_{x,y} = A e^{- B t} \sin(C t + D) + E ,
\end{equation}
where, in general, the undetermined parameters $A$-$E$ are independent between $P_x$ and $P_y$.  Note that as $t \rightarrow \infty$, $P \rightarrow E$, so we expect each fit for the parameter $E$ should agree with the equilibrium values (Eq.\ (\ref{eq:ringing_eqm})).  When we fit the solutions for $P_x$ and $P_y$, we find that the parameter $E$ in both fits are consistent with the equilibrium values.  Furthermore, for both $P_x$ and $P_y$, we find the the angular frequency, $C = \vert \v{V} \vert = \omega$, is equal to the in-medium neutrino oscillation rate, and the decay rate, $B = \Gamma /2$, is equal to half the active neutrino scattering rate.  

\begin{figure}
    \includegraphics[width=\columnwidth]{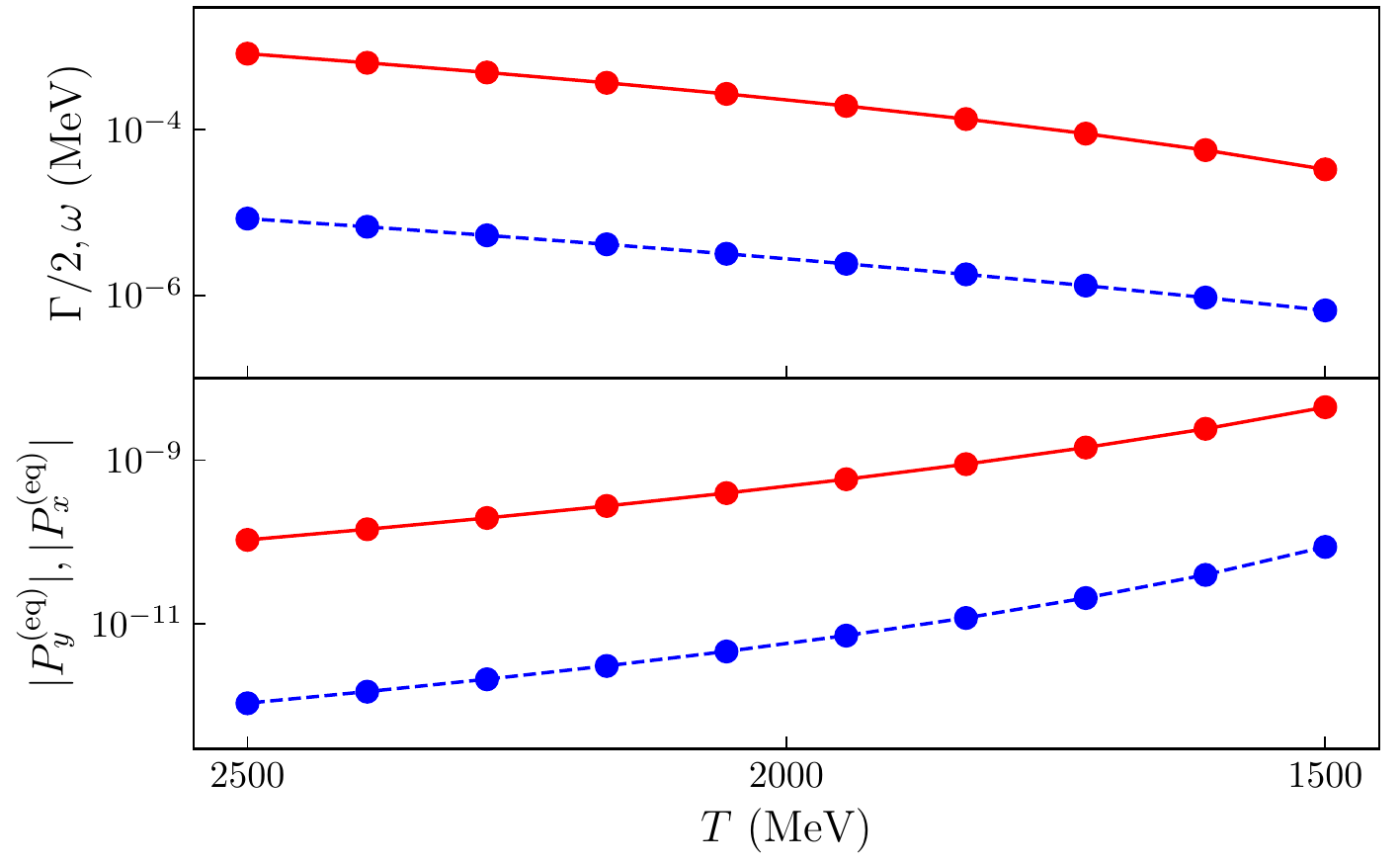}
    \caption{The QKEs are solved with initial conditions $\v{P} = \hat{\mathbf{z}}$, $P_0 = 1$, and different initial temperatures.  The subsequent ringing of $P_x$ and $P_y$ are fit to a damped exponential.  The results of this fit are shown as a function of initial temperature, with oscillation rate (top panel, solid curve), $\omega$, damping rate (top panel, dashed curve), $\Gamma / 2$, and equilibrium values of $P_x$ and $P_y$ (bottom panel, solid and dashed curves, respectively), $P_x^{\rm (eq)}$ and $P_y^{\rm (eq)}$.  Shown are analytic curves of the expected values (solid and dashed curves) and distinct data points that are the result of individual numerical experiments, run at different initial temperatures.} \label{fig:fits}
\end{figure}

Finally, we observed this ringing phenomenon from the initial conditions to an equilibrium at different initial temperatures.  With different initial temperatures, we numerically solved the QKEs and found a best fit for $P_x$ and $P_y$ with a damped oscillator model.  We find that in each scenario, $P_x$ and $P_y$ ring toward their equilibrium (as defined in Eq.\ (\ref{eq:ringing_eqm})) with oscillation rate $\omega$ and damping rate $\Gamma / 2$.  Figure \ref{fig:fits} shows the results of these fits at different initial temperatures.  The solid and dashed curves are the temperature-dependent values of $\omega$, $\Gamma / 2$, $P_x^{(\rm eq)}$, and $P_y^{(\rm eq)}$.  The dots are the best fit values from the fits as described above.  It is evident that the initial conditions ring toward equilibrium with oscillation rate, $\omega$, and damping rate, $\Gamma / 2$, independent of the initial temperature.

\section{Quantum Kinetic Equilibrium} \label{sec:qkeq}

\begin{figure}
    \includegraphics[width=\columnwidth]{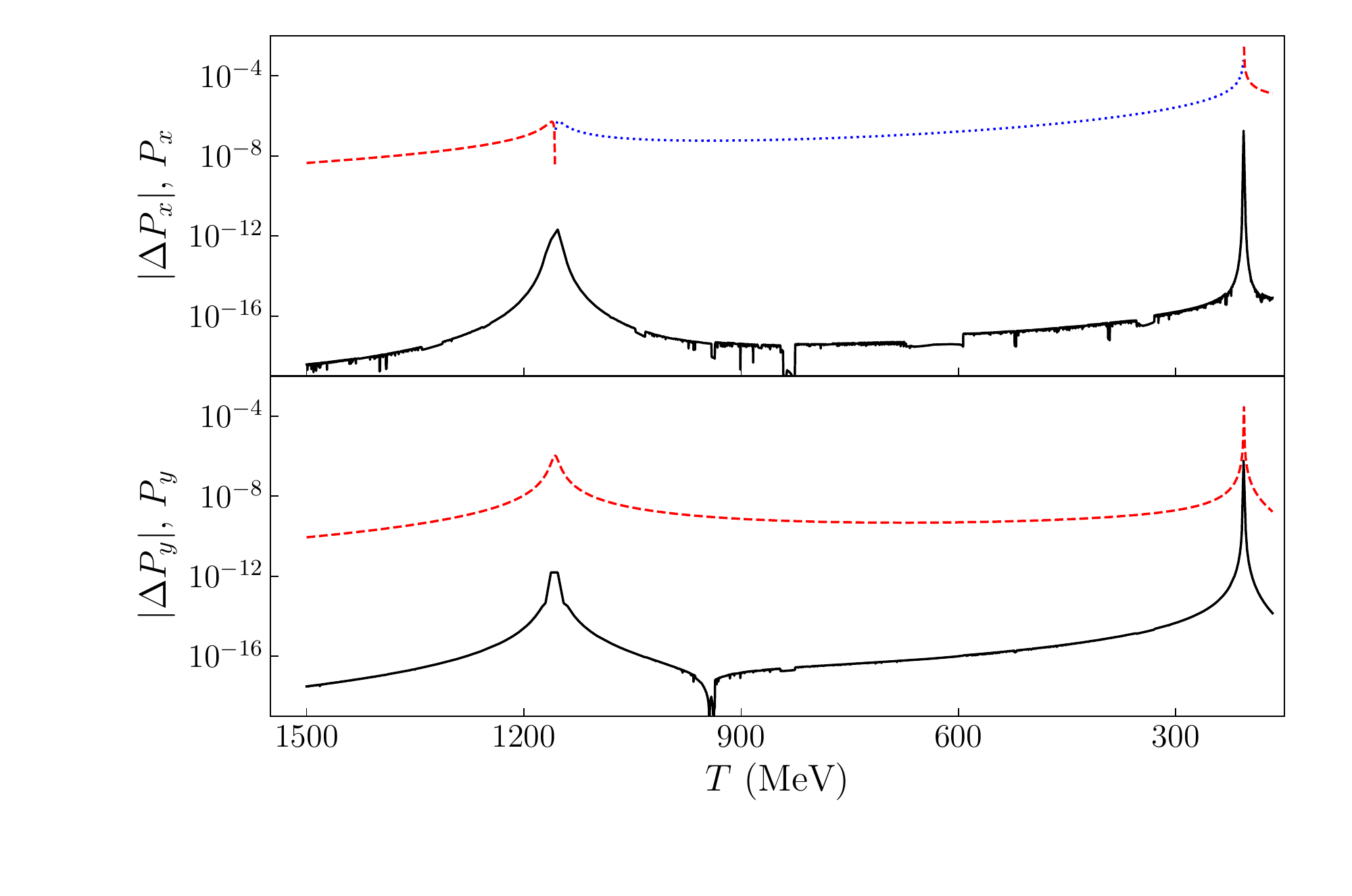}
    \caption{The QKEs are solved in the early universe with an initial lepton number as described in the introduction to mimic the production of a dark matter-candidate sterile neutrino.  In the figure, the solutions to the QKEs for $P_x$ and $P_y$ are shown as the upper curve in each panel.  The dotted portions (seen only in $P_x$) are positive values and the dashed are negative values.  Each panel also shows the difference between $P_x$ and $P_y$ and their equilibrium values as described in Eq.\ (\ref{eq:qkeqm_total}).  We solve the QKEs from an initial temperature of $1500~{\rm MeV}$, and the temperature decreases as the universe expands, so in the figures, time moves from left to right.} \label{fig:diff}
\end{figure}

The damped oscillations in $P_x$ and $P_y$ that ring from the initial conditions to the equilibrium progress with exponentially decreasing amplitude until there is no discernible oscillation, within the numerical precision.  The equilibrium values depend only on the plasma temperature; this temperature evolves [Eq.\ (\ref{eq:dTempdt})] as the universal expansion rate.  We note that at the temperatures of interest (e.g., the initial conditions studied in Fig.\ \ref{fig:fits}), both the oscillation rate, $\omega$, and decay rate, $\Gamma / 2$ of these damped oscillations are significantly fast compared to the expansion rate, $H$.  (In Figure \ref{fig:rates}, we see that they are $\gtrsim 10^9$ times faster than the expansion rate.)  As a result, we can effectively treat the temperature -- and hence the equilibrium values of $P_x$ and $P_y$ -- as constant during this entire process.  However, if we continue to evolve the QKEs from a higher initial temperature, we find that the polarization vector evolves along with the equilibrium values.

This observation leads us to a more general proposition:  whenever the oscillation and scattering rates are fast compared to the dynamical rate (the rate at which the properties of \v{V} evolves), the polarization vector evolves in a ``local'' quantum kinetic equilibrium where $dP_x/dt$ and $dP_y/dt$ are both approximately zero.  This local equilibrium evolves as \v{V} and $\Gamma$ evolves,
\begin{align}
    P_x^{\rm (eq)} & = \frac{\omega^2 \cos 2 \theta_M}{\omega^2 \cos^2 2 \theta_M + d^2} \sin 2 \theta_M P_z \nonumber \\
    P_y^{\rm (eq)} & = - \frac{\omega d}{\omega^2 \cos^2 2 \theta_M + d^2} \sin 2 \theta_M P_z ,
    \label{eq:qkeqm_total}
\end{align}
where $\omega = \vert \v{V} \vert$ is the in-medium neutrino oscillation frequency,
\[ \tan 2 \theta_M = \frac{V_x}{V_z} = \frac{\sin 2 \theta}{- \cos 2 \theta + \frac{2p}{m_s^2} V} \]
is the tangent of the angle between \v{V} and the +z-axis, and the overall damping rate of $\v{P}_\perp$ is
\begin{equation} 
    d = \frac{1}{2} \Gamma + \frac{1}{P_0} R .
\end{equation}
It should be noted that $R \propto \Gamma$, but is only significant fraction of $\Gamma$ near resonance and is otherwise very small.  In the limit that $P_0 = 1$ and $P_z = 1$ (and hence, $R = dP_0/dt = 0$) and $\theta_M \ll 1$ (and hence, $\cos 2 \theta_M \approx 1$, this result simplifies to the original equilibrium statement, Eq.\ (\ref{eq:ringing_eqm}).  However, this new proposition evolves as the universe expands which directly causes $T$ to decrease (and as a result, $\omega$ and $\Gamma$), but also allows for $P_z$ and $P_0$ to evolve.

Figure \ref{fig:diff} shows a comparison of the solutions to the QKEs with these equilibrium values.  In the figure, the solutions for $P_x$ and $P_y$ are shown with the dotted curve representing positive values of $P_x$ and the dashed part of the curves representing the negative values of $P_x$ and $P_y$. The solid curve on each plot shows the absolute value of the difference between the QKE solutions and these equilibrium values.  The QKEs were solved from initial conditions at $T = 1500~{\rm MeV}$ and allowed to evolve to a final temperature of 200 MeV, after the neutrinos have experienced two resonances -- level crossings -- where \v{V} changes rapidly.  These resonances occur at \mbox{$T_{\rm res} \approx 1160$} and $210~{\rm MeV}$ and can be seen on the figure where $P_x$ and $P_y$ change rapidly and where the difference between the actual solution and equilibrium solution is largest.  These two resonances and their concurrent rapid changes are depicted in parts (b) and (c) of Figure \ref{fig:QKE_solutions}.  At the resonance, the rapid change in \v{V} results in the rapid change of the proposed equilibrium state and $\omega$ decreases prodigiously (see, Figure \ref{fig:rates}).  So, it is, perhaps, unsurprising that the equilibrium approximation is most stressed at the resonances.

\begin{figure}
    \includegraphics[width=\columnwidth]{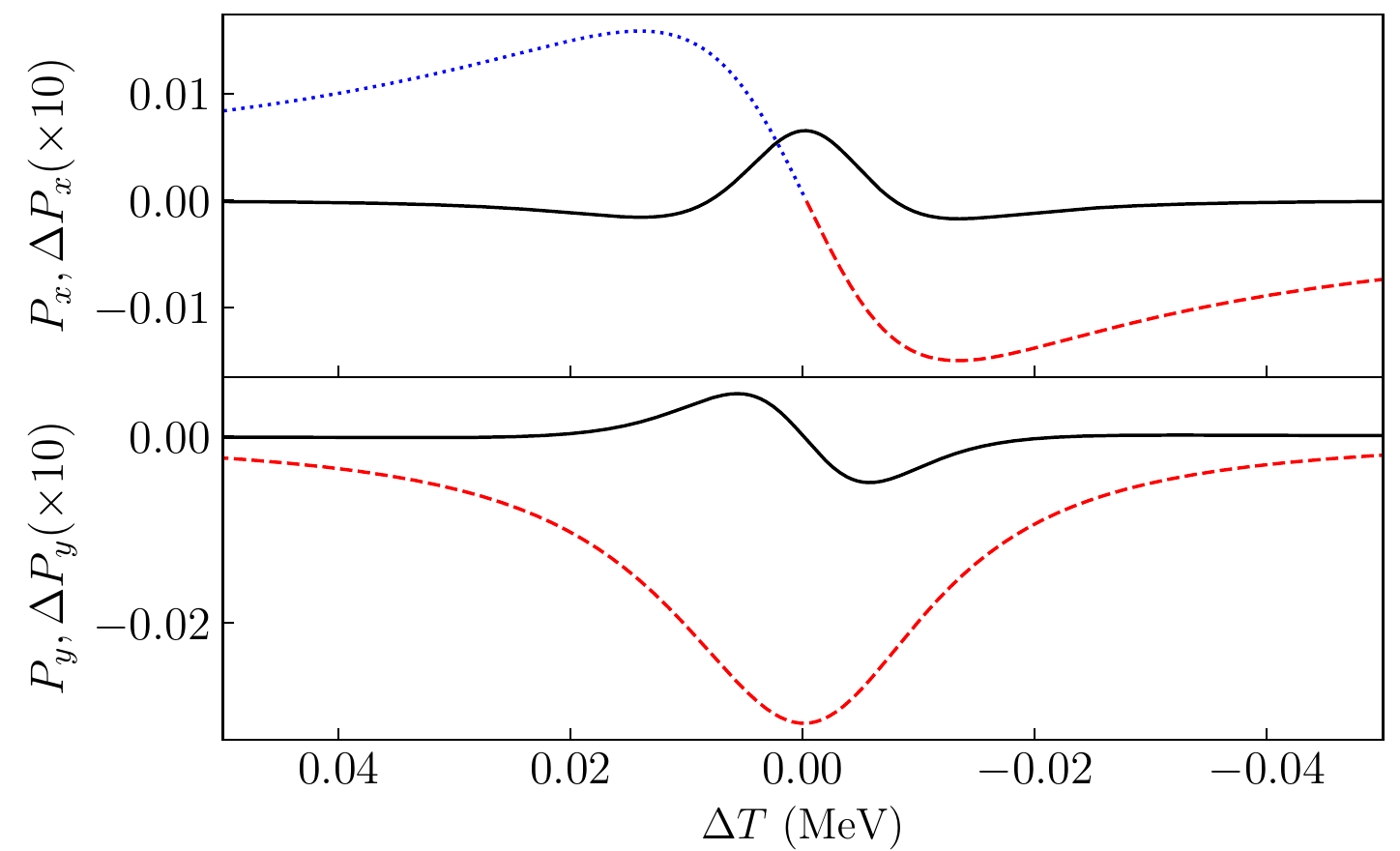}
    \caption{A zoomed-in view at the second resonance at \mbox{$T_{\rm res} \approx 206~{\rm MeV}$}, here in a linear view with $\Delta P_i \equiv P_i - P_i^{\rm (eq)}$, for $i = x, y$ and $\Delta T = T - T_{\rm res}$.  This figure uses the same scheme as Figure \ref{fig:diff}, with solid representing the $\Delta P_i$, the dotted curve is positive $P_x$, and the dashed curve are negative $P_i$.  The $\Delta P_i$ have been multiplied by 10 so that both curves can be viewed on the same plot.  As before, time flows from left to right in the figure as the temperature decreases in the expanding universe.} \label{fig:diffzoom}
\end{figure}

Far from resonance, we note that the quantum kinetic equilibrium, Eq.\ (\ref{eq:qkeqm_total}), is consistent with $P_x$ to approximately one part in $10^{10}$ and with $P_y$ to approximately one part in $10^8$.  (Compare this with the numerical tolerance of one part in $10^{10}$ that we used in the Cash-Karp adaptive step method.)  It appears that far from resonance, $\sin 2 \theta_M \ll 1$, the equilibrium approximation is a very good one.  Figure \ref{fig:diffzoom} shows a zoomed-in view of the second resonance at $T_{\rm res} \approx 206~{\rm MeV}$.  Immediately proximate to the resonance, we find that $P_x$ and $P_y$ deviate from the equilibrium values at the level of a few percent, at most.  It should be noted that the span represented in Figure \ref{fig:diffzoom} represents nearly 100 times the resonance width; so \v{V} flips almost instantaneously in this figure at $\Delta T = 0$.

\section{Discussion and Conclusions} \label{sec:conc}

We solved the quantum kinetic equations for an active-sterile neutrino state in the early universe in the presence of a net lepton number.  In this environment, the in-medium oscillation frequency and scattering rate is much greater than the rate at which the Hamiltonian evolves for nearly the entirety of the neutrino evolution.  Whenever this is true, we find that the polarization vector, \v{P}, achieves a local quantum kinetic equilibrium that slowly changes as the Hamiltonian changes.  This effective ``equilibrium'' attains as the unitary development of quantum coherence ({\it to wit}, quantum phase) is balanced by the de-coherent scattering-induced destruction of this coherence so that $d P_x/ dt \approx d P_y / d t \approx 0$, as defined by the equilibrium values of $P_x^{\rm (eq)}$ and $P_y^{\rm (eq)}$ as seen in Eq.\ (\ref{eq:qkeqm_total}).

\subsection{Dark Matter and the Quantum Zeno Ansatz}

The calculations include resonances where the Hamiltonian changes rapidly.  At these resonances, the in-medium oscillation frequency drastically decreases, which stretches the approximations made in the above statement of equilibrium.  However, when we examine the resonance, we find that the quantum kinetic equilibrium approximation continues to do an effective job at describing the quantum kinetic evolution of the polarization vector, albeit at the maximum discrepancy of a few percent.  The few percent discrepancy is notable when compared to past work in comparing solutions to the QKEs with the quantum Zeno ansatz for dark matter sterile neutrino production \cite{kf08}, which found percent-level discrepancies as well.  The quantum Zeno ansatz is an oft-used approximation for the scattering-induced decoherent production of sterile neutrino dark matter \cite{afp,kf08,vcah16,fv97,johns19} that incorporates an approximation of the quantum Zeno effect that suppresses resonant production (because of the decrease in the oscillation rate at resonance).

The quantum Zeno ansatz transforms the computationally expensive task of solving the quantum kinetic equations into the easier computational problem of solving Boltzmann-like equations.  With this ansatz, the Boltzmann equation for the sterile neutrino distribution function is
\begin{equation}
    \frac{d f_{ss}}{d t} = \frac{\Gamma}{4} \sin^2 2 \theta_M \frac{\omega^2}{\omega^2 + \frac{1}{4} \Gamma^2} (f_{ee} - f_{ss} ) .
\end{equation}
Since $f_{ss} = \frac{1}{2} P_0 (1 - P_z)$, we can use the QKEs and the equilibrium values for $P_x$ and $P_y$ to find a similar equation using this quantum kinetic equilibrium proposition,
\begin{equation}
    \frac{d f_{ss}}{d t} = \frac{d}{2} \sin^2 2 \theta_M \frac{\omega^2}{\omega^2 \cos^2 2 \theta_M + d^2} (f_{ee} - f_{ss}) .
\end{equation}
Far from equilibrium, $d \approx \Gamma / 2$ and $\cos^2 2 \theta_M \approx 1$, and the two agree with each other.  However, near the resonance, where both the quantum Zeno ansatz and this equilibrium have shown to be consistent with the full solutions to the QKEs, albeit at the level of a few percent.

When considering the possibility of sterile neutrinos as a dark matter candidate, it is important to understand not only the total production to be consistent with the total dark matter energy density from cosmological measurements, but to also discuss the momentum-distribution of such a dark matter distribution.  This spectrum is important to understand because it directly affects structure formation in the universe \cite{ch17}.  Ideally, one would solve the quantum kinetic equations for a distribution of neutrino momenta that are coupled primarily through neutrino-neutrino interactions in the forward scattering potential, but also through scattering.  This would represent a computationally expensive calculation that would inhibit parameter space searches for sterile neutrino dark matter properties.  However, this notion of a quantum kinetic equilibrium would drastically reduce the computational resources required to solve the remaining equations of motion.  In particular, by using $P_x^{\rm (eq)}$ and $P_y^{\rm (eq)}$ for the x- and y-components of the polarization vector away from equilibrium not only halves the number of ODEs to solve, but it also removes the fundamentally oscillatory nature of the QKEs that require tiny time steps in order to self-consistently solve these equations.  Furthermore, at resonance the neutrino in-medium oscillation frequency is at its minimum, so the largest step sizes in the entire calculation occur near resonance.  This means that if the notion of quantum kinetic equilibrium holds for the full calculation, its introduction would significantly and immensely reduce the computational cost of solving self consistently solving the QKEs for an ensemble of neutrino momenta.

\subsection{The Quantum Zeno Effect}

The quantum Zeno effect describes the suppression of quantum transitions at sufficiently high scattering rates.  We see in the equilibrium polarization vectors, $P_x^{\rm (eq)}$ and $P_y^{\rm (eq)}$, that when $\Gamma \gg \omega$, quantum coherences (represented by $P_x$ and $P_y$) are suppressed.  Further, transition probabilities are dictated by the evolution of $P_z$,
\begin{equation}
    \frac{d P_z}{d t} = V_x P_y^{\rm (eq)} + (1 - P_z) \frac{R}{P_0} .
\end{equation}
Which, through its dependence on $P_y^{\rm (eq)}$ shows that same suppression at large $\Gamma$.  This shows us that the quantum Zeno effect, as is typically advertised, is born of the suppression of quantum coherence caused by rapid scattering.  In a quantum kinetic sense, we see that the quantum Zeno effect is the result of the competition between the creation and destruction of quantum phase.  The greater the scattering rate, the smaller the x- and y-components of the polarization vector, and hence, the (in this case neutrino) state is very similar to one of the interaction eigenstates, suppressing the transition probability.

\subsection{Caveats and the Path Forward}
The many caveats of this calculation must first be addressed.  Two affects that occur on the relatively slow time scale of the expanding universe are the evolution of the relativistic degrees of freedom, $g_*$, and the evolution of lepton number as active neutrinos are converted to sterile ones.  We repeat this calculation with a time evolving $g_*$ \cite{kt90, ls06} and with time evolving lepton numbers from active-sterile neutrino dark matter transformation calculations using the above quantum Zeno ansatz \cite{kf08}.  In both cases, while the specifics of the shapes of the resonances change along with the total active-sterile neutrino transformation through the resonance, the quantum kinetic equilibrium approximation continues to do an excellent job of describing the actual solutions to the QKEs away from resonance and also shows a few-percent discrepancy at resonance.  As predicted, this form of time evolution occurs on a universal expansion time scale and since they is slow compared to the time scales that define neutrino oscillation and scattering,  they do not affect the general conclusions of this work.

The most important caveat, and the class of issues that are most ripe for future work, is the linearity of this solution.  While these solutions are valid to interpret linear situations, to solve neutrino-related quantum kinetic problems requires us to address the inherently non-linear problem.
So long as the oscillation and scattering rates are fast compared to the rate of change of the Hamiltonian (and the linearity of this solution demands that the evolution of the Hamiltonian is independent of the evolution of the neutrino density operators), the quantum kinetic equilibrium approximation remains a good one.  However, the forward scattering potential, Eq.\ (\ref{eq:V}), has the lepton numbers and neutrino energy densities which should reflect the feedback from the evolving density operators with
\begin{align}
    L &\propto \int d \epsilon \, \epsilon^2 \left[ P_0 (1 + P_z ) - \bar{P}_0  (1 + \bar{P}_z)  \right] ,\\
    \rho_\nu + \rho_{\bar\nu} & \propto \int d \epsilon \, \epsilon^3 \left[ P_0  (1 + P_z ) + \bar{P}_0 (1 + \bar{P}_z) \right] ,
\end{align}
where $\bar{P}$ represent components of the polarization vectors of anti-neutrinos (yes, this requires coupling with the anti-neutrino degrees of freedom), and the integrals incorporate and couple density operators of all $\epsilon$ values.  However, as these only affect $P_0$ and $P_z$, we expect initial consideration of this non-linearity response to be very similar to the previous experiment that used other, Boltzmann-like calculations to mimic the lepton number evolution.  While this could affect the details of the evolution (the temperature of the resonances and the total production of sterile neutrinos), we expect the general conclusions about the quantum kinetic equilibrium to be relatively unaffected.

However, this does not fully account for the neutrino-neutrino interactions, which includes both diagonal and off-diagonal interactions (terms both on and off the diagonal of the Hamiltonian, written as a matrix).  \cite{vve13}  The lepton-number driven term in the Hamiltonian (the first line of Eq.\ (\ref{eq:V}))  is proportional to
\begin{equation}
    \mathcal{H}_L \propto \int d \epsilon \, \epsilon^2 \left( f - \bar{f} \right) ,
\end{equation}
which when converted to the weak isospin space used here to solve the QKEs is
\begin{equation}
    \v{V}_L \propto \int d \epsilon \, \epsilon^2 \left( P_0  \v{P}  - \bar{P}_0  \v{\bar{P}} \right) .
\end{equation}
Likewise the so-called thermal term (the second line of Eq.\ (\ref{eq:V})) contains terms that are proportional to the energy density, which is proportional to 
\begin{equation}
    \mathcal{H}_T \propto \int d \epsilon \, \epsilon^3 \left( f + \bar{f} \right) ,
\end{equation}
which corresponds to
\begin{equation}
    \v{V}_T \propto \int d \epsilon \, \epsilon^3 \left( P_0 \v{P}  + \bar{P}_0  \v{\bar{P}}  \right) .  
\end{equation}

The upshot is that the neutrino-neutrino interactions inherent in both terms create both x- and y-components to the Hamiltonian vector, \v{V}, where non-linear feedback (integrated over all $\epsilon$) will affect the observed ringing phenomenon and the quantum kinetic equilibrium, as well as the possibility of introducing novel effects such as collective oscillations across ranges of $\epsilon$.  Solving this problem is a truly computationally expensive proposition, but understanding the characteristics of the solution is not only important in understanding the production of sterile neutrino dark matter, but also more broadly in the description of quantum kinetic phenomena.

\acknowledgements
We'd like to thank G.\ Fuller for useful discussions.  The work of CTK and HH is supported in part by NSF Grant PHY-1812383 and a Faculty Research Grant from the Dean of the College of Arts and Sciences at the University of San Diego.

\bibliography{nsf16.bib}
\end{document}